\documentclass[aps,prl,a4paper,superscriptaddress,twocolumn]{revtex4}
\usepackage{bm}
\usepackage{amsmath}
\usepackage{amsfonts}%
\usepackage{amssymb}%
\usepackage{graphicx}

\newcommand{\kt}{k_B T}

\begin{document}

\title{Multiscale modelling of liquids with molecular specificity}
\author{G. De Fabritiis}
\email[]{g.defabritiis@ucl.ac.uk}
\affiliation{Centre for Computational Science, Department of Chemistry, University College
London, 20 Gordon Street, WC1H 0AJ London, U.K.}
\author{R. ~Delgado-Buscalioni}
\email[]{rafa@ccia.uned.es}
\affiliation{Depto. Ciencias y T\'ecnicas Fisicoqu\'{\i}micas, 
Facultad de Ciencias, UNED, Paseo Senda del Rey 9,  Madrid 28040, Spain.}
\author{P. V. Coveney}
\email[]{p.v.coveney@ucl.ac.uk}
\affiliation{Centre for Computational Science, Department of Chemistry, University College
London, 20 Gordon Street, WC1H 0AJ London, U.K.}

\begin{abstract}
The separation between molecular  and mesoscopic length and time scales poses a
severe limit to  molecular simulations of mesoscale phenomena.  We describe a
hybrid multiscale computational technique which address this problem by keeping
the full molecular nature of the system where it is of interest and
coarse-graining it elsewhere. This is made possible by coupling  molecular
dynamics with a mesoscopic description of realistic liquids based on Landau's
fluctuating hydrodynamics.  We show that our scheme correctly couples
hydrodynamics and that fluctuations, at both the molecular and continuum levels,
are thermodynamically consistent.  Hybrid simulations of sound waves in bulk
water and reflected by a lipid monolayer are presented as illustrations of the
scheme. 
\end{abstract}
%\date{\today}
\maketitle

Complex multiscale phenomena are ubiquitous in nature in solid
(fracture propagation \cite{csa04}), gas (Knudsen layers \cite{Gar00})
or in liquid phases (fluid slippage past surfaces \cite{schmatko05},
crystal growth from fluid phase, wetting, membrane-fluid dynamics,
vibrational properties of proteins in water \cite{Tarek02,Baldini05}
and so on). 
These phenomena are driven by atomistic forces but manifest themselves at
larger, mesoscopic and macroscopic scales which cannot be resolved by purely 
large scale molecular simulations (with some notable exceptions 
\cite{Vashista96}). On the other hand, coarse-grained  mesoscopic models  have
limited use due to the approximations necessary to treat the molecular scales
intrinsic to these methods. A viable solution to this dilemma is represented by
multiscale modelling via coupled models, a protocol  which is also well suited
to new distributed computing paradigms such as Grids
\cite{foster05science,coveney05steering}.  The idea behind this approach is
simple: concurrent, coupled use of different physical descriptions.

The coupled paradigm is the underlying concept in
quantum-classical mechanics hybrid schemes \cite{csa04} used to describe fracture
propagation in brittle materials and also in hybrid models of gas flow
\cite{Gar00}. 
%RESUB: DELETED:::
%which couple shear, sound and energy transfer between a continuum
%fluid dynamics region and a rarefied subdomain solved by the direct simulation
%Monte Carlo technique. 
During the last decade, hybrid modelling of liquids has received
important contributions from several research groups (see the recent
review \cite{Kou05}). However, it has thus far
lacked the maturity to become a standard research tool for liquid and
soft condensed matter systems.  Hybrid simulations of liquids have
been restricted to coarse-grained descriptions based
on Lennard-Jones particles, reducing the major advantage of this
technique of maintaining full molecular specificity where
needed. Recently, new methods for energy controlled insertion of water
molecules \cite{ushers} have finally opened the way to real solvents
such as water.  So far, no hybrid method has employed an accurate
description of the mesoscale (from nanometres to micrometres) as the
important contribution of fluctuations has been neglected in the
embedding coarse-grained liquid. The hybrid method must
also ensure thermodynamic consistency, by allowing the open molecular
system to relax to an equilibrium state consistent with the
grand canonical ensemble \cite{flekbus}.  Finally, all previous
non-equilibrium hybrid simulations have been restricted to shear flow
\cite{Kou05,busF}.
%RESUB2  
%because of the difficulties in handling sound waves 
%which involve energy
%controlled fast molecule insertion into an open MD domain.

In this Letter, we present a coupled multiscale model called ``hybrid MD'' for
simulation of mesoscopic quantities of liquids (water) embedding a nanoscopic
molecular domain (Fig.  1a).  Hybrid MD overcomes the limitations
of previous hybrid descriptions of liquids by coupling  fluctuating
hydrodynamics \cite{landau59} and classical molecular dynamics via a 
protocol which guarantees mass and momentum conservation.  The
present method is designed to address phenomena driven by interplay between
the solute-solvent molecular interaction and the hydrodynamic flow of the solvent.

%In the following, we describe the FH model, the MD setting and the coupling
%protocol.  The model is validated under equilibrium and non-equilibrium
%conditions (shear and sound).  Finally we compare the hybrid model against full
%MD simulations of complex fluid flow: A  wave impinging against a lipid
%monolayer (dimyristoylphosphatidylcholine, DMPC).

{\em Fluctuating hydrodynamics model}.  Our mesoscopic description of
fluid flow is based on the equations of fluctuating hydrodynamics (FH)
\cite{landau59}.  These equations are stochastic partial differential
equations which reduce to the Navier-Stokes equations in the limit of
large volumes. The equations are based on the conservation equations
$\partial_{t}\phi =-\nabla \bm J^{\phi}$, where $\phi=\phi(\bm r,t)$
is the density of any conserved variable at location $\bm r$.  We
consider an isothermal fluid, so that the relevant variables are the
mass and momentum densities $\phi= \left\{\rho, \bm g\right\}$ (here
$\bm g \equiv \rho \bm v$ and $\bm v$ is the fluid velocity).  The
mass and momentum fluxes are given by $\bm J^{\rho}=\rho \bm v$ and
$\bm J^{\bm g}= \rho \bm v \bm v+ \bm \Pi+ \widetilde{\bm \Pi}$, where
$\bm \Pi$ and $\widetilde{\bm \Pi}$ are the mean and fluctuating
contributions to the pressure tensor, respectively. The mean pressure
tensor is usually decomposed as
$\mathbf{\Pi}=(p+\pi)\mathbf{1}+\mathbf{\Pi}^S$, where $p$ is the
thermodynamic pressure (given by the equation of state) and the stress
tensor is the sum of a traceless symmetric tensor $\bm \Pi^S$ and an
isotropic stress $\pi$.  We consider a Newtonian fluid for which $\bm
\Pi^S_{\alpha\beta} =- \eta \left( \partial_{\alpha}v_{\beta} +
\partial_{\beta}v_{\alpha} - 2 D^{-1}\partial_{\gamma}v_{\gamma}
\delta_{\alpha\beta} \right)$, $\pi = - \zeta \partial_{\gamma}
v_{\gamma}$, where repeated indices are summed, $D$ the spatial dimension  and $\eta$, $\zeta$ are
the shear and bulk viscosities respectively.
%RESUB::: DELETE (we state it later)
%(given for water at $T=300$K in \cite{visc_water}). 
The components of the fluctuating
pressure tensor $\widetilde{\Pi}_{\alpha\beta}$ are random Gaussian
numbers (see supplementary information).

Our continuum mesoscopic model is based on a finite volume discretization of
the FH equations \cite{serrano01}, although here in an Eulerian frame
of reference and on a regular lattice. Partitioning the space into several
space-filling volumes $V_{k}$ with $k=1,...,N$ centered at positions $\bm r_k$,
we integrate the conservation equations over each volume $V_{k}$ and apply
Gauss' theorem 
$\frac{d}{dt}\int_{V_{k}}\phi(\mathbf{r}_k,t)d \mathbf{r}=
%%%%\int_{V_{k}}\partial_{t}\phi(x,t)dx=
\sum\limits_{l} A_{kl} \mathbf{J}^{\phi}_ {kl}\cdot \mathbf{e}_{kl}$,
where $\mathbf{e}_{kl}$ is the unit surface vector  pointing towards cell $k$,
and  $A_{kl}$ is the surface area connecting cells $k$ and $l$.
We then derive the following stochastic equations for mass and momentum exchange:
\begin{align}
dM_{k}^{t}  & =
\sum_l \mathbf{g}_{kl} \cdot\mathbf{e}_{kl}A_{kl}dt,\label{massupdate}\\
d\mathbf{P}_{k}^{t}  & =\sum_l \left [\frac{\bm \Pi_{l}}{2}\cdot\mathbf{e}_{kl} 
+ \mathbf{g}_{kl} \cdot \mathbf{e}_{kl} \mathbf{v}_{kl}\right] A_{kl}dt 
+d\widetilde{\mathbf{P}}_{k}^{t},\label{momupdate}
\end{align}
where $d\widetilde{\mathbf{P}_k}$ is the momentum exchange 
due to the fluctuating pressure tensor $\widetilde{\Pi}_k$, 
${\bf v}_{kl}={\bf v}_k-{\bf v}_l$ and $\mathbf{g}_{kl}$ is approximated on the surface kl by 
$\mathbf{g}_{kl}=\frac{1}{2}(\rho_{k}+\rho_{l})\frac{1}{2}(\mathbf{v}_{k} + \mathbf{v}_{l})$.
To close the discrete conservation equations we have to devise a
discretization of the dissipative and fluctuating parts which ensures 
the validity of the fluctuation-dissipation theorem.  
By choosing the discretization of the gradients 
$\partial^{\alpha} \phi_k \rightarrow \sum_l A_{kl}\bm{e}^{\alpha}_{kl} \phi_k/(2V_k)$,
 the discrete momentum fluxes $\mathbf{\Pi}_{k}$ and 
 $d\widetilde{\mathbf{P}}_k$ take the form  given in \cite{serrano01} (see also supplementary information). 
The resulting set of stochastic differential equations
Eqs. (\ref{massupdate},\ref{momupdate}) 
may be integrated using various stochastic integration schemes \cite{defabritiis06_trotter}; in this work we have used a simple Euler scheme.

{\em Molecular dynamics.} The molecular description  is based on classical
molecular dynamics and the CHARMM27 forcefield (incorporating the TIP3P parametrization) which specifies bond, angle,
dihedral and improper bonded interactions and non-bonded Lennard-Jones 6-12 and
Coulomb interactions.  The code is derived from a stripped down version of NAMD
\cite{NAMD}. We use a dissipative particle dynamics (DPD) thermostat
\cite{soddemann03} ensuring local momentum conservation in
such a way that hydrodynamic modes are not destroyed.

\begin{figure}[tb]
\begin{center}
\includegraphics[width=7cm]{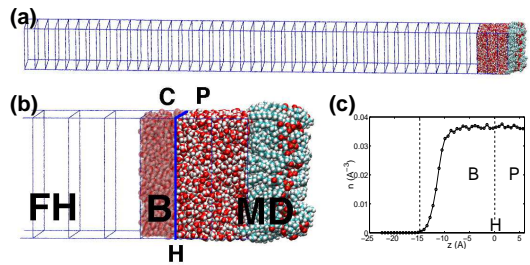}
\caption{The set-up used for our hybrid molecular simulations (a) and a close-up of
the hybrid interface (b). The fluctuating
hydrodynamics description (FH), resolved by the finite volume method, is
coupled to a molecular model (MD) representing a dimyristoylphosphatidylcholine
(DMPC) lipid monolayer solvated with water and  restrained at the lipid
head groups.  We indicate by ``P'' and ``C'' respectively the particle and continuum
cells adjacent to the hybrid interface ``H''.  The buffer region of the MD
system ``B'' (overlapping the C cell) is indicated by translucent water
molecules and the water molecule density in the buffer region is shown in (c).}
%\label{hybridfig}
\end{center}
\end{figure}

%-------------------------------------------------------------------------
{\em Coupling protocol.-}  In our computational implementation,
the MD and FH components are independent {\em coupled models} \cite{coveney05steering} 
which exchange information after every fixed time interval $\Delta t_{c}$.
We set $\Delta t_{c}=n_{FH} \Delta t= n_{MD}\delta t$, where
$\Delta t$ and $\delta t$ are the FH and MD time steps and, $n_{FH}$
and $n_{MD}$ are integers which depend on the system being modeled;
e.g. for water as solvent $\Delta t_c = 100$ fs, $n_{FH}=10$ and
$n_{MD}=100$. Conservation is based on the flux balance: both
domains receive equal but opposite mass and momentum fluxes across the
hybrid interface. This interface (H) uniquely defines the total system
(MD+FH, see Fig. 1b) and, importantly, the total quantities to be
conserved. This contrasts with previous schemes
\cite{Kou05} where particle and continuum domains intertwine within a larger
overlapping region, preventing a clear definition of the system.

The rate of momentum transferred across the hybrid interface is given
by ${\bf F}_H= A \bm J^{\bm g}_{H} \cdot \bm e_{\perp}$, where $\bm
e_{\perp}$ is the unit vector perpendicular to the surface and the
momentum flux tensor at ``H'' is approximated as $\bm J^{\bm g}_{H} =
(\bm J^{\bm g}_P+\bm J^{\bm g}_C)/2$.  Note that $\bm J^{\bm g}_C$
involves the evaluation of the discretized velocity gradient at $C$,
and thus requires the mass and momentum of the MD system
at the neighbouring P cell averaged over the coupling time $\Delta
t_C$: $\langle M_P \rangle_{\Delta t_c}$ and $\langle \bm P_P \rangle_{\Delta t_c}$, respectively (see Fig. 1b).
%Here $M_P=\sum_i \chi_P({\bf r}_i)m_i$ and
%${\bf P}_P=\sum_i \chi_P({\bf r}_i)m_i {\bf v}_i$ ere $\chi_P$
%is the characteristic function of the cell P, i.e. $\chi_P({\bf r})=1$ if ${\bf
%r} \in P $ and $0$ otherwise.  
On the other hand, the momentum flux tensor at the P cell can be
computed for the microstate using the kinetic theory formula $\bm
J^{\bm g}_P = \langle [\rho {\bf v}_i{\bf v}_i+W_i]\rangle_{\Delta t_c}$, with $i\in P$ and
$W_i=\sum_j {\bf r}_{ij} \cdot {\bf f}_{ij}$ \cite{allen87} being
the contribution of atom $i$ to the virial.  Alternatively, $\bm
J^{\bm g}_P$ can be computed by introducing the coarse-grained variables at the
neighboring MD and FH cells into the discretized Newtonian
constitutive relation.  Both approaches provide equivalent results in terms
of mean and variance of the pressure tensor.

The force ${\bf F}_H$ at the hybrid interface is imposed on the FH
domain using standard von Neumann boundary conditions. In order to
impose the force $-{\bf F}_H$ on the molecular system, we extend the
MD domain to an extra buffer cell (``B'' in Fig. 1b).  Particles are
free to cross the hybrid interface according to their local dynamics,
but any atom that enters in B will experience an external force 
$-{\bf F}_H/N_B$ which transfers the external pressure and stress.  The
number of solvent  molecules at the buffer $N_B(t)$ is controlled by a simple
relaxation algorithm: $\Delta{N}_B=(\langle N_B\rangle -N_B)\Delta
t_c/\tau_B$, with $\tau_B\simeq 500$ fs.  The average $\langle N_B
\rangle$ is set so as to ensure that B always contains enough
molecules to support the momentum transfer;
%RESUB
here we use $\langle N_B \rangle = 0.75 M_C/m$, where $M_C$ is the
mass of the continuum cell C and $m$ the molecular mass.
%%RESUB
Figure 1c shows the equilibrium number density profile of
water at the buffer.  Importantly, the density profile is flat around
the hybrid interface.  Due to the external pressure, it quickly
vanishes near the open boundary.  In fact, molecules eventually
reaching this rarefied region in B are removed. If the relaxation
equation requires $\Delta N_B >0$, new molecules are placed in B with
velocities drawn from a Maxwellian distribution with mean equal to
the velocity at the C cell.  The insertion location is determined by
the {\sc usher} algorithm \cite{ushers}, which efficiently finds new
molecule configurations releasing an energy equal to the mean energy
per molecule.  Momentum exchange due to molecule insertion/removal is  
taken into account in the overall momentum balance \cite{flekbus}.

In fluid dynamics the mass flux is not an independent quantity but is
controlled by the momentum flux [see Eqs. (\ref{massupdate}) and
(\ref{momupdate})]. Consequently, we do not explicitly impose the mass flux on the MD
system.  Instead it arises naturally from the effect of the external
pressure on the molecule dynamics near the interface. 
The mass flux $\bm J^{\rho}_H\cdot \bm e_{\perp}$ is
thus measured (via simple molecule count) from the amount of MD mass 
crossing the interface H over the coupling time $\Delta t_c$. The
opposite of this flux is then transfered to the adjacent C cell via a
simple relaxation algorithm \cite{flekbus}, using a relaxation time
($\tau_r \ge O(100)$ fs) large enough to preserve the correct
mass distribution at the C cell, but still much faster than any
hydrodynamic time. This guarantees mass conservation. 

%-------------------------------------------------------------------------

{\em Results.} We first test the conservation of the total mass
$M$ and momentum ${\bm P}$. Results are shown
in Fig. 2a, where we consider the equilibrium state of
a hybrid MD simulation of water in a 3D periodic box
$50\times 50\times 735$ $ \mathrm{\AA}^3$ (each cell is $50\times 50\times 15$ $ \mathrm{\AA}^3$).  
The embedded TIP3P water domain
(including the buffers) is $75 \mathrm{\AA}$ wide in the coupling (z) direction
and was pre-equilibrated at 1 atm and 300K.  Figure 2a shows the mean
error in mass and momentum conservation. As stated above, mass
conservation is ensured over a short time $\Delta t_c \sim O(100)$ fs,
as clearly reflected in Fig. 2a. However, as the external force is
imposed within the buffers B, the momentum conservation is ensured
only on the ``extended'' system (MD+FH+B).  The variation of momentum of the total
system (MD+FH) is then a small bounded quantity whose time
average becomes smaller than the thermal noise after about
1ps (see Fig. 2a), i.e, faster than any hydrodynamic time scale.

\begin{figure}[tb]
\begin{center}
\includegraphics[width=4cm, height=3.5cm]{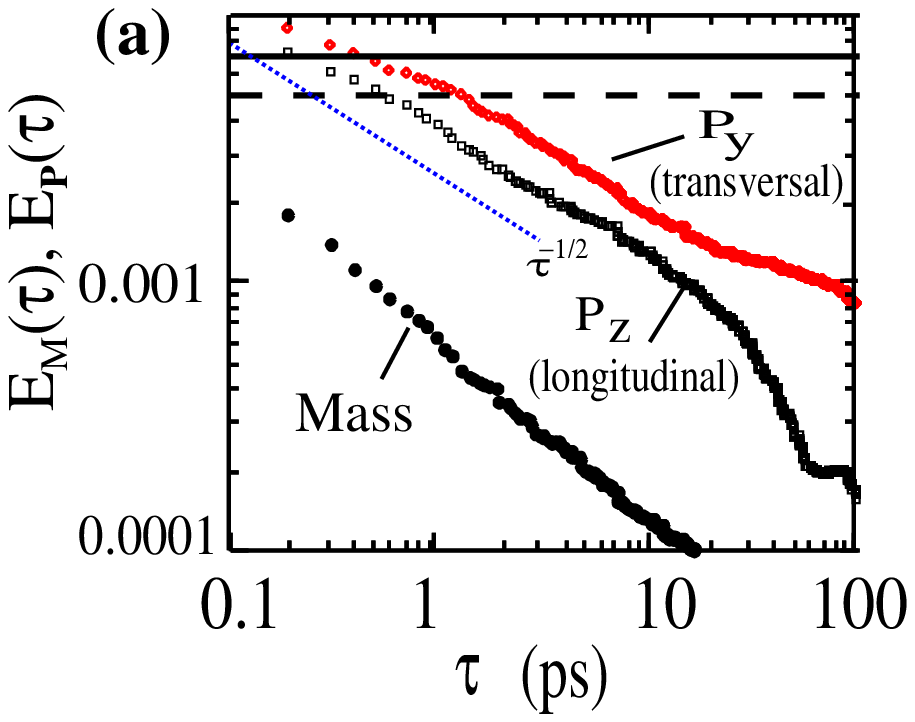}
\includegraphics[width=4cm, height=3.5cm]{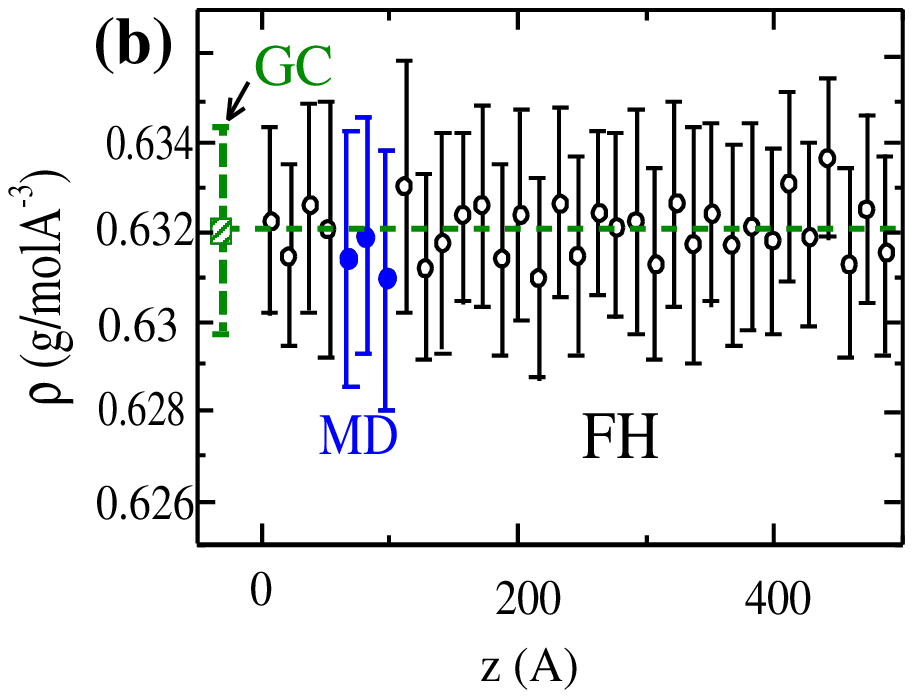}
\caption{(a) The normalized mean error in mass $E_M(\tau)$ and
momentum $E_{P_{\alpha}}(\tau)$ evaluated as $E_A^2(\tau)=\langle
[\int_{t_0}^{t_0+\tau} \delta A(t) dt/\tau ]^{2} \rangle_{t_0}/\langle
M_k\rangle^2$, with $\delta A=A-\langle A\rangle$. The dashed and
solid horizontal lines are, respectively, the normalized standard
deviation of mass and momentum within one cell ($\sigma[M_k]/\langle
M_k\rangle$).  (b) Density field in a hybrid MD equilibrium simulation
of water. Solid circles corresponds to MD cells.  Error bars are the
standard deviation of each cell density. The grand-canonical (GC)
result is $\langle \rho \rangle =0.632 \mathrm{g/mol/\AA^3}$ and
$\sigma[\rho]=0.0045 \mathrm{g/mol/\AA^3}$.  }
\end{center}
\end{figure}
The FH description uses an  accurate interpolated  equation of state
 $p(\rho)=(3.84-15.7\rho + 15.3\rho^2)\,10^4$
bars, which fits for $\rho=[0.54,0.70]$ $\mathrm{g/mol/A^3}$ the outcome of $NPT$ simulations of
TIP3P water at $T=300$K and provides quasi-perfect match of the mean
pressure, density (see Fig. 2b) and sound velocity.
%%RESUB
The shear and bulk viscosities of the FH model are assigned to match
those of the MD fluid (for water at $T=300$K we used those reported in
Ref. \cite{visc_water}). Also, in cases where the viscosity varies locally,
the FH model allows one to assign a different viscosity for each cell.
Momentum fluctuations at each cell are consistently controlled by the
DPD thermostat in the MD region, and via the fluctuation-dissipation
balance in the FH domain. Density fluctuations present a much more
stringent test of thermodynamic consistency.
Each fluid cell is an open subsystem so, at equilibrium, its mass fluctuation
should be governed by the grand canonical (GC) prescription:
$\sigma[\rho]=[\rho \kt/(V_k c_T^2)]^{1/2}$ \cite{landau59} (where $\sigma$
means standard deviation and $c_T^2\equiv (\partial P/\partial \rho)_T$ is the
squared sound velocity at constant temperature). Mass fluctuations within the
MD and FH cells are both in agreement with the GC result (Fig.
2b) indicating that neither the {\sc usher} molecule insertions
\cite{ushers} nor the mass relaxation algorithm
substantially alter the local equilibrium around the interface H.
%resub2
%(similar thermodynamic consistency is obtained for more compressible fluids like argon).

We now focus on transmission of sound waves  which thus far have
remained an open problem in the hybrid setting. 
%RESUB
In a slot of water between rigid walls we perturb the equilibrium
state with a Gaussian density perturbation (amplitude 5\% and standard
deviation $45 \mathrm{\AA}$).  As shown in Fig. 3a the resulting
travelling waves cross the MD domain several times at the center of
the slot.  Sound waves require fast mass and momentum transfer as any
significant imbalance would generate unphysical reflection at the
hybrid interface. No trace of reflection is observed and comparison
with full FH simulations shows statistically indistinguishable
results.

\begin{figure}[t!]
\begin{center}
\includegraphics[width=4cm,height=3cm]{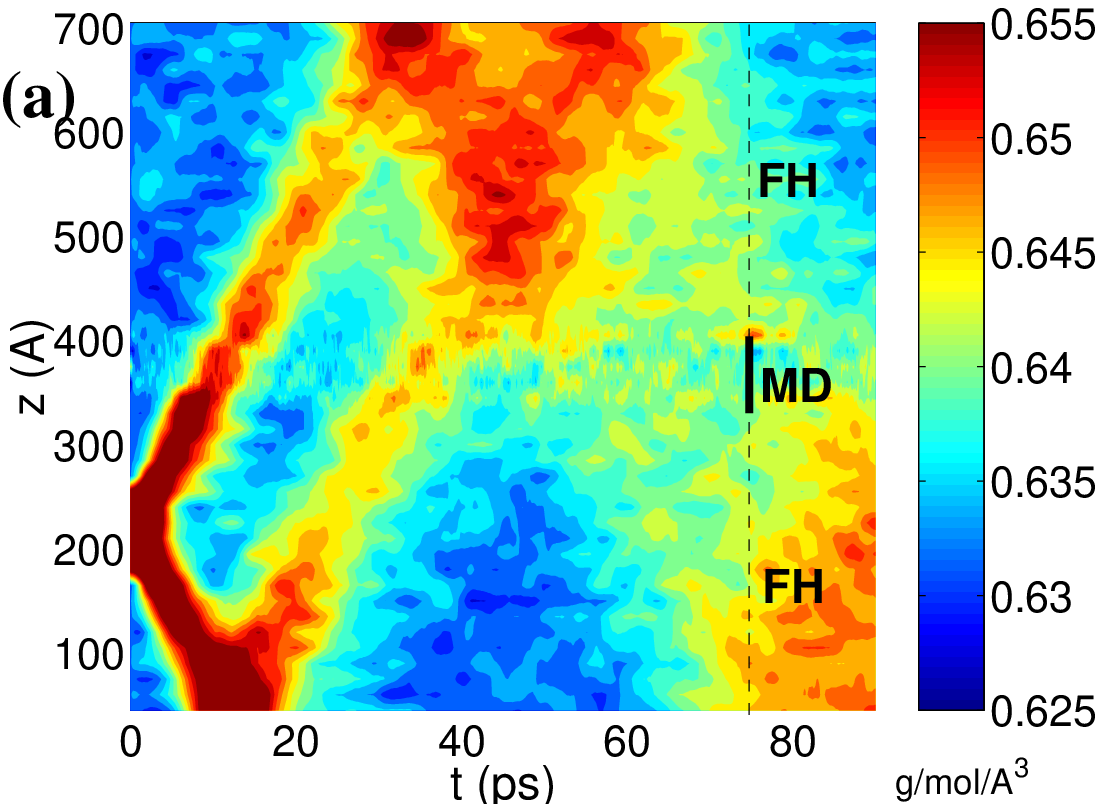}
\includegraphics[width=4cm,height=3cm]{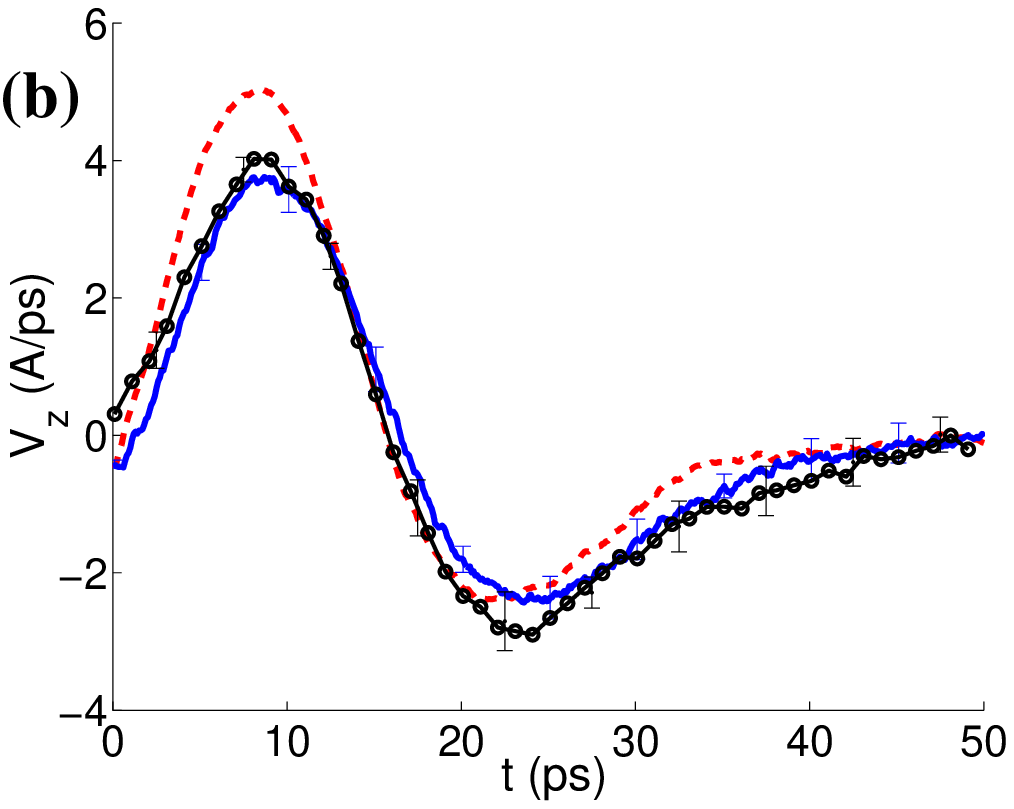}
\caption{(a) Spatio-temporal diagram along $z$ for the density field 
 of a three-dimensional simulation of two sound waves traveling within a closed box filled
with water.  The region of width $45 \mathrm{{\AA}}$ around the centre of the box
is described with molecular dynamics (MD), while the rest of the domain is
solved via fluctuating hydrodynamics (FH).
(b) The longitudinal velocity arising from the interaction 
between a sound wave and a grafted lipid layer (set up of Fig. 1a).
We compare hybrid MD (solid line), full MD (circles) and
full FH simulation using purely reflecting walls (dashed line). 
Results are averaged over $15$ nm from
the monolayer; error bars indicate the standard deviation over 10 runs.}
\end{center}
\end{figure}

Finally, we validate the hybrid scheme against full MD
simulations of complex fluid flow (set-up Fig. 1a). 
A sound wave generated by a similar Gaussian perturbation
is now reflected against a lipid monolayer (DMPC)  (Fig. 3b). 
Each lipid is tethered by the heavy
atoms of the polar head group with an equilibrated grafting cross-section of  
53 $\mathrm{{\AA}}^2$/lipid, close to the experimental cross-section of  
membranes. In the hybrid simulation, the MD water layer close to the
lipid membrane extends just $45 \mathrm{\AA}$ above it (see
Fig. 1b). Instead, in the MD simulation we considered a
large $180\times50\times50$ $ \mathrm{\AA}^3$  box of explicit water containing around 50K atoms.
%RESUB (details of velocity already in caption)
The wave velocity near the layer is compared in Fig. 3b 
for the hybrid MD and MD simulations. 
The excellent agreement demonstrates that the coupling
protocol accurately resolves features produced by the molecular structure.
%within the hybrid region. 
In Fig. 3b  such effects are due to 
sound absorption by the lipid layer, highlighted by comparison with
a FH simulation of the same wave impinging against a purely reflecting
wall.
%RESUB
The present sound waves simulations were
done assuming an isothermal environment.  This is realistic 
if the rate of thermal relaxation $D_T k^2$ (with $D_T\sim 1.5\times
10^{-7} m/s^{2}$ the water thermal diffusivity and $k=2\pi/\lambda$ the
wavenumber) is comparable with or faster than its sound frequency $c\, k$.  
The present simulations $\lambda\sim 50 \AA$ are just in the
limit of the isothermal sound regime \cite{Cowan96}, while
waves with $\lambda >O(10)\AA$ propagate adiabatically and require
consideration of the energy flow \cite{flekbus}.

%RESUB2 Only for length checking purposes...anyhow, if some parragraph 
%goes out this is a good candidate in my opinion

%The reduction in computational cost is roughly equal to the ratio of
%atoms in the hybrid and MD simulations, being around one order of
%magnitude for the tests of Fig. 3b.  Here, longer
%wavelengths are easily tackled by extending the FH domain at
%negligible cost but, by contrast, increasing a standard MD simulation box
%becomes rapidly prohibitive.

%resub2 (i've included stable and robust at the beginning)
In summary, we have presented a stable and robust
multiscale method (hybrid MD) for the
simulation of the liquid phase which embeds a small region, fully
described by chemically accurate molecular dynamics, into a
fluctuating hydrodynamics representation of the surrounding
liquid. Mean values and fluctuations across the interface are consistent with
hydrodynamics and thermodynamics.  Sound waves propagating through the
MD domain and flow behavior arising from the interaction with complex
molecules are both treated correctly. We considered water waves reflected by
DMPC monolayers, but the scope of this methodology is much broader,
including {\em inter alia} the study of vibrational properties of
hydrated proteins (via high frequency perturbations)
\cite{Tarek02,Baldini05}, the ultrasound absorption of complex liquids
\cite{usound} or the simulation of quartz crystal oscillators 
\cite{broughton97} for the study of complex fluid rheology 
or slip flow past surfaces \cite{schmatko05}.

%{\footnotesize GDF\&PVC are grateful to EPSRC for funding the Integrative Biology project GR/S72023
%and BBSRC for funding the IntBioSim project BBS/B/16011.
%RD-B acknowledges projects MERG-CT-2004-006316, CTQ2004-05706/BQU and
%FIS2004-01934.  
%We thank M. Serrano, A. Dejoan, S. Succi, 
%P. Espa{\~n}ol and E. Flekk{\o}y for various discussions and assistance.
%}
{\footnotesize GDF\&PVC acknowledge projects Integrative Biology (GR/S72023)
and IntBioSim (BBS/B/16011).
RD-B acknowledges projects MERG-CT-2004-006316, CTQ2004-05706/BQU and
FIS2004-01934.  
We thank M. Serrano, A. Dejoan, S. Succi, 
P. Espa{\~n}ol and E. Flekk{\o}y.% for various discussions and assistance.
}
\end{document}